\documentclass[sigplan,10pt]{acmart}
\renewcommand\footnotetextcopyrightpermission[1]{}
\usepackage{graphicx} 
\usepackage{subcaption} 
\usepackage{xspace} 
\usepackage{acronym} 
\usepackage[dvipsnames]{xcolor}
\usepackage{tikz}
\usepackage{enumitem}

\newcommand{\sys}{Nanvix\xspace}
\newcommand{\sysShort}{NVX}

\newcommand{\systemVm}{system VM\xspace}
\newcommand{\systemVms}{system VMs\xspace}
\newcommand{\SystemVm}{System VM\xspace}
\newcommand{\userVm}{user VM\xspace}
\newcommand{\userVms}{user VMs\xspace}
\newcommand{\UserVm}{User VM\xspace}
\newcommand{\UserVms}{User VMs\xspace}
\newcommand{\multiKernel}{multikernel\xspace}
\newcommand{\smallKernel}{micro-kernel\xspace}
\newcommand{\smallVMM}{VMM\xspace}
\newcommand{\bigKernel}{macro-kernel\xspace}
\newcommand{\SmallKernel}{Micro-kernel\xspace}
\newcommand{\BigKernel}{Macro-kernel\xspace}

\makeatletter
\renewcommand{\sectionautorefname}{\S\@gobble}
\renewcommand{\subsectionautorefname}{\S\@gobble}
\renewcommand{\subsubsectionautorefname}{\S\@gobble}

\makeatother

\newcommand{\eg}{e.g.\xspace}
\newcommand{\ie}{i.e.\xspace}
\newcommand{\code}[1]{{\texttt{#1}}\xspace}

\newcommand{\myparagraph}[1]{\vspace{\smallskipamount}\noindent\textbf{#1.\xspace}}

\newcommand*\myColourCirc[3]{%
  \scalebox{0.78}{%
    \begin{tikzpicture}[baseline=-3pt]
      \node[
        circle,
        inner sep=0.5pt,
        fill=#1
      ]{\textcolor{#2}{\textsf{\textbf{#3}}}};
    \end{tikzpicture}%
  }%
}

\definecolor{MyBlue}{HTML}{0072B2} 
\definecolor{MyOrange}{HTML}{E69F00} 
\definecolor{MySkyBlue}{HTML}{56B4E9}
\definecolor{MyYellow}{HTML}{F0E442}
\definecolor{MyRed}{HTML}{D55E00} 
\definecolor{MyBluishGreen}{HTML}{009E73}
\definecolor{MyPink}{HTML}{CC79A7} 
\definecolor{MyGray}{HTML}{999999}

\acrodef{vm}[VM]{virtual machine}
\newcommand{\vm}{\ac{vm}\xspace}
\newcommand{\vms}{\acp{vm}\xspace}

\acrodef{os}[OS]{operating system}
\newcommand{\os}{\ac{os}\xspace}

\acrodef{cow}[CoW]{copy-on-write}
\newcommand{\cow}{\ac{cow}\xspace}

\acrodef{ksm}[KSM]{Kernel Samepage Merging}
\newcommand{\ksm}{\ac{ksm}\xspace}

\acrodef{slo}[SLO]{service level objective}
\newcommand{\slo}{\ac{slo}\xspace}

\acrodef{pci}[PCI]{Peripheral Component Interconnect}

\begin{document}

\title{\sys{}: A Multikernel OS Design for High-Density Serverless Deployments}

\author{
  \rm
    Carlos Segarra$^{\dagger}$ \enskip
    Pedro Henrique Penna$^{\ddagger}$ \enskip
    Enrique Saurez$^{\star}$ \enskip
    \'{I}\~{n}igo Goiri$^{\star}$ \enskip
    \\ \rm
    Peter Pietzuch$^{\dagger}$ \enskip
    Shan Lu$^{\ddagger\triangle}$ \enskip
    Rodrigo Fonseca$^{\star}$ \enskip
  \\
  {
    \small
    $^{\dagger}$Imperial College London
    $^{\ddagger}$Microsoft Research
    $^{\triangle}$University of Chicago
    $^{\star}$Azure Research - Systems
  }
}

\begin{abstract}
    Serverless providers strive for high resource utilization by optimizing \emph{deployment density}: how many applications can be deployed per host server. However, achieving high deployment density without compromising application performance or isolation remains an open challenge. High density can be achieved by sharing components across applications, yet applications from different tenants must be strongly isolated from each other due to the risk of side-channel attacks. Sharing components across applications from the same tenant, if done naively, can introduce contention on host resources thus negatively affecting application performance.

    We describe \emph{\sys{}}, a new \multiKernel OS that disaggregates ephemeral execution state, unique per application invocation, from long-lived persistent state, shared among invocations from the same tenant. Applications in \sys{} execute inside a lightweight \emph{\userVm} running a \smallKernel that implements threads and memory, and forwards all I/O requests to a \emph{\systemVm}. The \systemVm runs a \bigKernel with a rich set of device drivers and is shared among all invocations from the same tenant. \sys{}' split design achieves strong hypervisor isolation across tenants without sacrificing application performance, and reduces same-tenant contention by multiplexing all I/O requests to the \systemVm. Thanks to a system-wide co-design, \sys{} achieves order-of-magnitude lower application start up times with moderate I/O overheads. When replaying a production trace, \sys{} needs 20-100$\times$ fewer host servers compared to state-of-the-art systems, improving deployment density.
\end{abstract}

\settopmatter{printfolios=true}
\maketitle
\pagestyle{plain}


\section{Introduction}
\label{section:introduction}

The past decade of serverless research has driven significant reductions in application cold start latency through lightweight isolation (\eg micro-VMs)~\cite{Firecracker:NSDI:2020,Virtines:EuroSys:2022}, fast snapshot restoration~\cite{Catalyzer:ASPLOS:2020}, elastic scaling under bursty load~\cite{AFaaS:OSDI:2025,Dandelion:SOSP:2025}, and increasingly sophisticated orchestration layers~\cite{Dirigent:SOSP:2024}. As a result, production serverless systems can launch new applications in less than a hundred milliseconds, even in periods of high contention~\cite{RunD-V:TOCS:2025,AFaaS:OSDI:2025}.

Cold start latency, however, is  \emph{not} the only metric that needs to be optimized. From the perspective of the serverless provider, equally important are \emph{inter-tenant isolation} and \emph{deployment density} -- how many applications can be co-located on a single server while still meeting per-request latency \slo targets~\cite{RunD-V:TOCS:2025}. Unfortunately, it is hard to achieve deployment density while maintaining low cold start latencies and strong isolation: high density can be achieved by sharing components across applications, but excessive sharing can introduce resource contention and side-channels. An orthogonal but equally desirable property is compatibility with existing programming frameworks: being able to run existing software on standard runtimes, instead of requiring a completely new programming model and targeted rewrites~\cite{Dandelion:SOSP:2025,Faasm:ATC:2020,Virtines:EuroSys:2022}.

\begin{figure}
    \centering
    \includegraphics[width=\columnwidth]{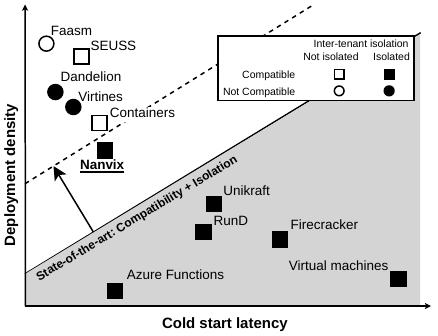}
    \caption{\textbf{Serverless design space} (Serverless providers wish to optimize deployment density as a proxy for resource efficiency, but struggle to do so while maintaining low cold start latencies, strong inter-tenant isolation, and application compatibility.)}
    \label{figure:introduction:trilemma}
\end{figure}

\autoref{figure:introduction:trilemma} illustrates the trade-offs made by different systems. Many prior works relax inter-tenant isolation by sharing the same process substrate via software-fault isolation~\cite{Faasm:ATC:2020, CloudFlareWorkers:2026}, or the same \os kernel, via containerization~\cite{Fission:2026, SEUSS:EuroSys:2020}. In this paper, we target multi-tenant serverless deployments that cannot compromise on inter-tenant isolation and must isolate applications using \vms.

Systems such as Azure Functions, explicitly sacrifice deployment density for lower cold start latency by pre-warming \vm instances in a pool~\cite{DROPS:EuroSys:2026}. Other systems achieve deployment density, cold start latency, and inter-tenant isolation, but sacrifice application compatibility by imposing substantially different programming and execution models~\cite{Dandelion:SOSP:2025, Virtines:EuroSys:2022}. Unfortunately, the systems that do retain compatibility (under virtualization)~\cite{Firecracker:NSDI:2020,Unikraft:EuroSys:2021,RunD-V:TOCS:2025} end up paying a large, often order-of-magnitude, density and latency penalty compared to those which do not. This is because they use a guest \os that was not designed for ephemeral execution and work around it with advanced \vm snapshot techniques~\cite{AFaaS:OSDI:2025,AwsSnapStart:2026} or using lightweight isolation inside a parent \vm~\cite{RunD-V:TOCS:2025}. Ultimately, these system-level solutions shift the pressure to other parts of the system (see \autoref{section:motivation}).

\begin{figure}[t]
    \centering
    \begin{subfigure}{.49\linewidth}
        \includegraphics[width=\linewidth]{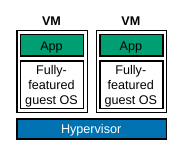}
        \caption{Traditional \vms}
        \label{figure:introduction:overview:vm}
    \end{subfigure}
    ~
    \begin{subfigure}{.49\linewidth}
        \includegraphics[width=\linewidth]{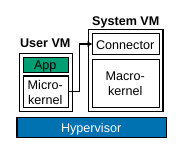}
        \caption{\sys}
        \label{figure:introduction:overview:multikernel}
    \end{subfigure}

    \caption{\textbf{\sys{} is a \multiKernel \os for serverless} (Existing multi-tenant serverless platforms offer compatibility by including all \os functionality in the guest,  fundamentally limiting performance and density. \sys{}, on the other hand, advocates for a disaggregated design.)}
    \label{figure:introduction:overview}
\end{figure}

On this landscape, our key observation is that the high cost of compatibility is the root cause of cold start latency and low deployment density. This cost cannot go down in a secure and fundamental way as long as we expect the guest \os inside each \vm to implement all the features required for compatibility such as devices, network and file systems (see \autoref{figure:introduction:overview:vm}). Consequently, to achieve a minimum virtualization environment for serverless applications, we argue for \emph{disaggregating} the ephemeral execution state of every application from the long-lived persistent state of the \os.

\myparagraph{Our proposal}
To fulfill this idea, we present \emph{\sys{}}, a new \multiKernel \os for high-density serverless deployments. \sys{} splits the system services of the guest \os across two virtualized environments: (1) a lightweight \emph{\userVm} with a \smallKernel that supports application's ephemeral execution with thread and memory management; (2) a feature-rich \emph{\systemVm} with a \bigKernel that provides system functionality (\ie networking and file system) for all \userVms belonging to the same tenant (see \autoref{figure:introduction:overview:multikernel}).

\sys is a unique \multiKernel \os where two kernels provide complementary properties that together meet serverless performance, compatibility, security, and density goals:
\begin{description}
    \item[\SmallKernel.]
    It implements processes, threads, scheduling, memory management, IPC, and a POSIX compatibility layer, but eschews device virtualization, network, or file systems. This minimal-feature design ensures each \userVm has a small memory footprint and very-fast startup, addressing the cold start latency without compromising deployment density or security.
    \item[\BigKernel.]
    It includes the full set of device drivers and kernel modules needed to serve I/O requests that the \smallKernel cannot handle. The \systemVm, similar to other cloud \vms running a fully-featured \os, has a considerably higher cold start latency and memory footprint than a \userVm. Fortunately, unlike ephemeral \userVms, a \systemVm is long-lived, serves requests from many \userVms from the same tenant, and can be proactively auto-scaled. In addition, given that it does not execute tenant-specific code, all \systemVm instances can be restored from a generic snapshot.
\end{description}

We adopt a clean-slate (co-)design for the \smallKernel, its \smallVMM, and the userspace components of the \systemVm, while employing a trimmed Linux as the \bigKernel. We implement all components in Rust and release them as open-source. \sys{} executes unmodified applications written in C, C++ and Rust after compilation with our toolchain. It also executes Python, JavaScript, and WebAssembly~(WASM) applications by compiling the language runtime.

Our evaluation shows that \userVms start up in a handful of milliseconds, an order-of-magnitude faster than restoring a micro-\vm from a snapshot~\cite{REAP:ASPLOS:2021}. A \systemVm and a \userVm are collectively deployed in less than 30 milliseconds. Thanks to the explicit sharing of the \systemVm, we can deploy 30--50\% more sandboxes per GiB than other snapshot-based baselines. When replaying a production trace, a system using \sys would require 20--100$\times$ fewer servers to serve the trace compared to one using Firecracker~\cite{Firecracker:NSDI:2020} or optimized snapshot-based systems.

\section{Towards Dense Serverless Deployments}
\label{section:motivation}

We next argue why deployment density is an important metric for serverless providers~(\autoref{subsection:motivation:beyond-cold-start}), how can it be achieved securely~(\autoref{subsection:motivation:sharing}), and why is it challenging to do so~(\autoref{subsection:motivation:challenges}).

\subsection{Why deployment density?}
\label{subsection:motivation:beyond-cold-start}

Early workload characterizations of production serverless deployments showed that cold start time dominated request latencies~\cite{PeekingCurtains:ATC:2018,ServerlessInTheWild:ATC:2020,ServerlessColdStartsAndWhereToFindThem:EuroSys:2025}. Since then, serverless research has focused on reducing cold start latencies by optimizing every component on the request critical path, from the control-plane~\cite{Dirigent:SOSP:2024} to the execution sandbox~\cite{Firecracker:NSDI:2020}. However, cold start latency, as a proxy for request latency, is a \emph{tenant-facing} metric and does not directly measure serverless provider's efficiency. Some production serverless systems report function cold starts below 100~ms~\cite{AFaaS:OSDI:2025}, while others offer getting rid of cold starts altogether for a price premium~\cite{AwsProvisionedConcurrency:2026,DROPS:EuroSys:2026}.

\emph{Deployment density}, on the other hand, is a provider-facing metric. Achieving high deployment density means using fewer physical servers to serve the same load, yielding higher utilization, lower cost, and higher revenue from providers' perspective. Deployment density is determined by the dominant resource. If the workload consumes a lot of memory, such as using large container images or deploying large pre-warmed \vm pools~\cite{DROPS:EuroSys:2026}, memory capacity will dominate deployment density. On the other hand, if the workload follows a bursty arrival pattern, which is common in serverless~\cite{DROPS:EuroSys:2026,ServerlessInTheWild:ATC:2020}, application start throughput, which is often dominated by contention on host \os resources~\cite{RunD-V:TOCS:2025,AFaaS:OSDI:2025}, can limit the deployment density.

Note that these two metrics are related. Techniques that shorten the cold start latency can reduce the amount of time each request needs to get served, and hence may improve the deployment density through increased time sharing among requests on each server. However, if the cold start latency is improved by consuming extra system resources, the deployment density may drop as discussed in \autoref{section:introduction}.

\subsection{How to improve deployment density?}
\label{subsection:motivation:sharing}

A common approach to improving deployment density is to share resources across applications. However, in a multi-tenant serverless environment, there is little room for inter-tenant sharing due to many security constraints. The hypervisor cannot allocate applications from different tenants in sibling cores that share micro-architectural state nor employ memory sharing techniques like \ksm due to the risk of side-channel attacks~\cite{DedupEstMachina:SP:2016,Meltdown:Security:2018,Spectre:SP:2019,MemDedup:DSN:2013}. Snapshot images may be shared, but under specific circumstances that we expand on in \autoref{subsection:motivation:challenges}.

\begin{figure}[t]
    \centering
    \includegraphics[width=\linewidth]{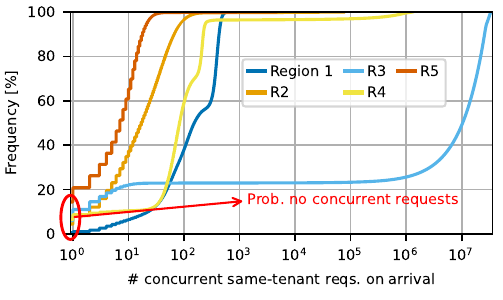}
    \caption{\textbf{CDF of the \# of in-flight requests from the same tenant when a request arrives}
    (We plot the CDF across five different regions based on the Huawei trace from 2025~\cite{ServerlessColdStartsAndWhereToFindThem:EuroSys:2025}. In red we highlight the frequency of there being \emph{zero} in-flight requests from the same tenant.)}
    \label{figure:motivation:amortization}
\end{figure}

Intra-tenant sharing is therefore the focus of this paper. In \autoref{figure:motivation:amortization} we show that, even without inter-tenant sharing, there is a big opportunity to share resources across applications of the same tenant. Based on the Huawei trace from 2025~\cite{ServerlessColdStartsAndWhereToFindThem:EuroSys:2025}, there is an 80\% probability that when a request arrives there is another in-flight request from the same tenant, creating resource sharing opportunities.

\subsection{Why is sharing hard?}
\label{subsection:motivation:challenges}

Existing approaches to exploit intra-tenant sharing are constrained by using a full guest \os to offer application compatibility, and need to work around this limitation to offer high deployment density.

\myparagraph{Containers in \vms}
A common approach is to deploy same-tenant applications inside containers, instead of \vms, inside a parent, per-tenant, \emph{base \vm}. This deployment strategy amortizes the cost of the guest \os, image layers, and system libraries across application invocations of the same tenant~\cite{RunD-V:TOCS:2025}. However, it introduces a fundamental sizing tension.

\begin{figure}[t]
    \centering
    \begin{subfigure}{.48\linewidth}
        \includegraphics[width=\linewidth]{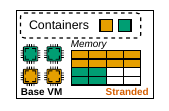}
        \caption{Static base \vm}
        \label{figure:motivation:ctrs-in-vms:static}
    \end{subfigure}
    ~
    \begin{subfigure}{.5\linewidth}
        \includegraphics[width=\linewidth]{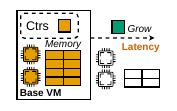}
        \caption{Dynamic base \vm}
        \label{figure:motivation:ctrs-in-vms:dynamic}
    \end{subfigure}

    \caption{\textbf{Containers in \vm deployment} (This approach employs a base \vm that can either have a static allocation of resources, in which case some resources may be stranded, or may grow and shrink dynamically, adding latency to the request's critical path.)}
    \label{figure:motivation:ctrs-in-vms}
\end{figure}

\begin{table}[t]
    \centering
    \small
    \setlength{\tabcolsep}{3pt}
    \begin{tabular}{lccccccccc}
    \toprule
        vCPU & 1 & 4 & 8 & -- & -- & -- & 1 & 4 & 8 \\
        Mem (MiB) & -- & -- & -- & 128 & 512 & 1024 & 128 & 512 & 1024 \\
    \midrule
        \textbf{p50 (ms)} & 73.5 & 94 & 141 & 22 & 22.5 & 25 & 77 & 86 & 141 \\
        \textbf{p99 (ms)} & 104 & 226 & 490 & 24 & 24.9 & 36.1 & 89.5 & 108 & 380 \\
    \bottomrule
    \end{tabular}
    \caption{\textbf{Hotplug latency} (We measure the time to hotplug vCPU and/or memory with ACPI and \code{virtio-mem}, respectively, and report p50 and p99.)}
    \label{table:motivation:hotplug}
\end{table}

\autoref{figure:motivation:ctrs-in-vms} shows two approaches to deploying this containers-in-\vms scheme. When the base \vm has a fixed allocation of vCPUs and memory (see \autoref{figure:motivation:ctrs-in-vms:static}), the scheduling of requests to base \vms boils down to a bin-packing problem. This approach is likely to strand resources if requests cannot be perfectly bin-packed or base \vms become fragmented over time. To avoid such resource stranding, we can allow the base \vm to grow or shrink dynamically based on request load (see \autoref{figure:motivation:ctrs-in-vms:dynamic}). However, if the growing or shrinking happens proactively, as part of an auto-scaling policy, it will often leave underutilized resources~\cite{MeldingControlPlane:arXiv:2026}, impacting density. If it happens reactively, the time to grow or shrink the base \vm contributes to the cold start latency. In \autoref{table:motivation:hotplug}, we show that hot-plugging a vCPU or memory balloon adds over 80 ms of overhead, and hot-unplugging can be substantially worse~\cite{RunD-V:TOCS:2025,Squeezy:EuroSys:2026}.

\myparagraph{\vm snapshots}
An alternative to avoid the resource stranding and performance overheads of containers in \vms is to deploy each application in a different \vm restored from a snapshot. What memory contents are included in the snapshot and how they are restored determines whether the same snapshot can be used to restore \vms from different tenants. For a snapshot to be safely re-used across tenants it must include neither tenant code nor system state that may have been influenced by the execution of tenant code like the GC~\cite{FrozenGarbage:EuroSys:2024} or JIT~\cite{Pronghorn:EuroSys:2024} compiler state in a managed language runtime such as Python, Java, or .NET. In addition, the guest \os must reinitialize the entropy pool and other sources of randomness after a restore~\cite{FirecrackerRandomForClones:2026,HandlingUniquenessSnapStart:2026}.

In practice, AWS Lambda SnapStart~\cite{AwsSnapStart:2026} and Ant FaaS~\cite{AFaaS:OSDI:2025} use per-function and per-tenant snapshots, respectively. To address the increased disk and memory pressure of having more and diverse snapshots, each system employs different advanced snapshot management mechanisms. SnapStart chunks snapshots in blocks and serves them on-demand from a distributed storage layer with two levels of caching across availability zones. Ant FaaS uses a hierarchical tree storing incremental snapshots restored with \cow. There is no quantitative analysis on the impact of these techniques on the host server's memory and disk pressure in production, but they undeniably add considerable management and system complexity.

\myparagraph{Summary}
Deployment density is an efficiency metric that directly correlates with provider's resource utilization. It is related to, but different from, cold start latency. Existing multi-tenant systems struggle to achieve high density and low cold start latency due to the reliance on heavyweight guest OSs to guarantee application compatibility.

\section{\sys Overview}
\label{section:system}

Our goal in \sys{} is to execute serverless applications in a minimum virtual execution environment while retaining application compatibility. Our approach, as highlighted in \autoref{figure:introduction:overview:multikernel}, is to disaggregate the system services offered by the guest \os in those essential for the execution of applications and those necessary to interact with external devices, and serve both from different virtualized environments that can communicate with each other.

\begin{figure}[t]
    \centering
    \includegraphics[width=.9\linewidth]{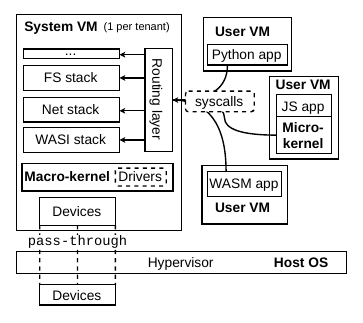}
    \caption{\textbf{Overview of \sys} (\sys adopts a \multiKernel design where a \smallKernel executes application code and a \bigKernel multiplexes I/O requests from \userVms belonging to the same tenant.)}
    \label{figure:system:overview}
\end{figure}

\sys{} adopts a \multiKernel~\cite{Barrelfish:SOSP:2009} design where a \smallKernel executes serverless applications and forwards all I/O system calls to a \bigKernel serving I/O requests for all applications from the same tenant (see \autoref{figure:system:overview}). Both kernels and their supporting systems software are co-designed for the specific purpose of executing short-lived ephemeral applications, and this co-design enables \sys{} to achieve all the design goals from \autoref{figure:introduction:trilemma}.

\myparagraph{Low cold start latency}
With high probability (see \autoref{figure:motivation:amortization}), when a new invocation request arrives there is already a running \systemVm instance for that tenant, so only the \userVm start up is on the critical path. The \userVm emulates a very simple machine model and thus avoids expensive initialization operations like scanning ACPI tables or enumerating PCI devices. The micro-kernel boots in one or two milliseconds, and the end-to-end latency for a cold HTTP echo takes a handful of milliseconds, including the control-plane (see \autoref{table:evaluation:micro:cold-start}).

\myparagraph{High deployment density}
Each application executes in a different \userVm and all I/O is multiplexed by a single \systemVm (per tenant). This disaggregation addresses the resource stranding issues of containers-in-\vms and reduces resource contention because each \userVm gets its own share of vCPU and memory, and only \systemVms contend for expensive network resources like namespaces or TAP devices. In addition, all \systemVms are restored from the same generic snapshot, further reducing \sys{} memory footprint and improving deployment density.

\myparagraph{Inter-tenant isolation}
The only shared substrate among \userVms hosting application from different tenants is the underlying hypervisor. On the other hand, \userVms that host applications from the same tenant share  the \systemVm. Worker threads in the \systemVm executing system calls from different \userVms from the same tenant can be further isolated from each other. In addition, \sys leverages the communication layer between the \systemVm and the \userVm as a low-friction vantage point for system call filtering and interposition (see \autoref{figure:system-vm:overview}).

All-in-all, \sys intra-tenant isolation guarantees are comparable to those of containers that share the container runtime (\eg containerd) and guest \os, used in production serverless systems like Google's CloudRun~\cite{GoogleCloudRun:2026}, Alibaba Serverless Containers~\cite{RunD-V:TOCS:2025}, and other Kubernetes or Knative based systems~\cite{Fission:2026}. Systems like Azure Functions and AWS Lambda also re-use the same container instance (and thus underlying \os state) through the function's \code{keep-alive} configuration~\cite{AwsLambdaRuntimeEnvironment:2026}.

\myparagraph{Application compatibility}
The cost of specialization and co-design is compatibility. The \userVm can only execute guests running the \sys \smallKernel. Instead of offering kernel-level compatibility, we aim to offer source code level compatibility by equipping the \smallKernel with a POSIX compatibility layer that enables the execution of unmodified applications after re-compilation. This type of compatibility is a common assumption in other serverless systems~\cite{Unikraft:EuroSys:2021, Junction:NSDI:2024}.

\vspace{5pt}

\sys{}' disaggregated design presented in \autoref{figure:system:overview} favors \emph{modularity}. The filesystem or network stack can be inside the \bigKernel, in userspace~\cite{Junction:NSDI:2024}, or entirely offloaded to hardware~\cite{AzureRDMA:NSDI:2023}, and stacks can be replaced transparently to application code. In addition, other type of API calls, not only I/O system calls, can also be forwarded from the \userVm to the \systemVm. For example, \sys support for WASM applications forwards most WASI~\cite{Wasi:2026} calls to the \systemVm. In a cloud environment where POSIX or system calls are less frequent~\cite{SigmaOS:SOSP:2024,Dandelion:SOSP:2025}, \userVms could also forward other cloud-native APIs like S3 to the \systemVm, or directly to the storage node.

\section{\UserVm}
\label{section:user-vm}

The \emph{\userVm}~(\autoref{subsection:user-vm:overview}) is \sys{}' virtual execution environment. It comprises a \smallKernel~(\autoref{subsection:user-vm:kernel}) and a \smallVMM~(\autoref{subsection:user-vm:vmm}) co-designed for low start-up times and near-zero-copy I/O.

\subsection{Overview}
\label{subsection:user-vm:overview}

\begin{figure}[t]
    \centering
    \includegraphics[width=.9\linewidth]{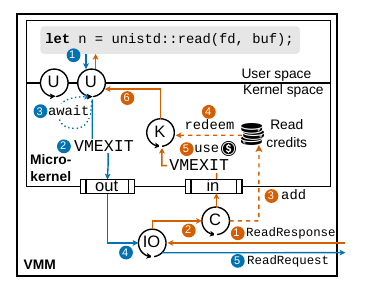}
    \caption{\textbf{\UserVm architecture} (The \userVm consists of a \smallKernel and a \smallVMM co-designed for fast start-up times and to minimize \vm exits during synchronization and I/O.)}
    \label{figure:user-vm:overview}
\end{figure}

\autoref{figure:user-vm:overview}~provides a high-level overview of the \userVm architecture, contrasting it with state-of-the-art VMMs capable of hosting Linux~\cite{Firecracker:NSDI:2020,CloudHypervisor:2026}, and illustrates the process of executing a \code{read} POSIX call. Application code in \sys is executed by one or many User threads in the \smallKernel~(\myColourCirc{MyBlue}{white}{1}, \autoref{figure:user-vm:overview}). After transitioning to kernel space and issuing the (remote) blocking call via port-mapped I/O~(PMIO)~\myColourCirc{MyBlue}{white}{2}, the user thread waits for the response~\myColourCirc{MyBlue}{white}{3}. The \smallVMM has a dedicated I/O thread~\myColourCirc{MyBlue}{white}{4} that forwards requests to the \systemVm~\myColourCirc{MyBlue}{white}{5} (not depicted). The same I/O thread receives the response and forwards it to the \smallKernel~\myColourCirc{MyRed}{white}{1}.

As part of the \userVm co-design, the \smallVMM-to-\smallKernel data path uses a credit-based flow control to minimize the number of \vm exits issued by the guest \smallKernel while polling for the response~\myColourCirc{MyRed}{white}{2}. A dedicated Credits thread adds a credit to a control page in the guest \smallKernel via shared memory~\myColourCirc{MyRed}{white}{3}. A dedicated Kernel thread monitors the credit counter and, when credits are available~\myColourCirc{MyRed}{white}{4}, issues a VM exit~\myColourCirc{MyRed}{white}{5} and forwards the response to the user thread~\myColourCirc{MyRed}{white}{6}. We expand on the credit-based control-flow as well as our design for bulk data transfers in \autoref{subsection:user-vm:vmm}.

\subsection{\SmallKernel}
\label{subsection:user-vm:kernel}

\begin{table*}[t]
    \centering
    \footnotesize
    \begin{tabular}{p{5cm} p{12cm}}
    \toprule
        \multicolumn{1}{c}{\textbf{Category}} & \multicolumn{1}{c}{\textbf{Kernel calls}} \\
    \midrule
        \multicolumn{2}{c}{\textit{Local execution}} \\
    \hline
    \textbf{Process \& Thread management} &
    \texttt{getpid}, \texttt{gettid}, \texttt{exit}, \texttt{exit\_thread}, \texttt{join\_thread}, \texttt{create\_thread} \\

    \textbf{Scheduling \& Synchronization} &
    \texttt{yield}, \texttt{sleep}, \texttt{mutex\_lock}, \texttt{mutex\_unlock}, \texttt{cond\_wait}, \texttt{cond\_signal}, \texttt{resume} \\

    \textbf{Memory management} &
    \texttt{mmap}, \texttt{munmap}, \texttt{mctrl}, \texttt{mcopy} \\

    \textbf{Capability \& Process control} &
    \texttt{capctl}, \texttt{terminate} \\

    \textbf{Thread-local storage} &
    \texttt{set\_thread\_data\_area}, \texttt{get\_thread\_data\_area} \\

    \textbf{Time \& Debug} &
    \texttt{gettime}, \texttt{debug} \\

    \textbf{Device I/O management} &
    \texttt{mmio\_alloc}, \texttt{mmio\_free}, \texttt{mmio\_info}, \texttt{pmio\_alloc}, \texttt{pmio\_free}, \texttt{pmio\_read}, \texttt{pmio\_write} \\

    \hline
        \multicolumn{2}{c}{\textit{Remote execution}} \\
    \hline

    \textbf{Inter-kernel communication} &
    \texttt{send}, \texttt{recv}, \texttt{push}, \texttt{pull} \\

    \bottomrule
    \end{tabular}
    \caption{\textbf{Classification of kernel calls by functionality and execution locality} (Most kernel calls are resolved locally in the \smallKernel, while communication primitives are forwarded to the \bigKernel in the \systemVm for remote execution.)}
    \label{table:user-vm:kcalls}
\end{table*}

\autoref{table:user-vm:kcalls} summarizes the different kernel calls implemented in the \userVm's \smallKernel. The POSIX compatibility layer in userspace translates POSIX calls, \eg~\code{unistd::read}, into the corresponding kernel call. The \smallKernel implements threads, processes, and memory management, including thread-local storage. The \smallKernel also implements thread scheduling and synchronization, capability-based access control, and memory-mapped I/O~(MMIO) and PMIO. All of these kernel calls are executed locally, within the kernel, and do not rely on the \systemVm.

The \smallKernel also implements a set of inter-kernel (\ie \smallKernel to \bigKernel) communication kernel calls. \code{send/recv} use regular PMIO to send one-off messages, whereas \code{push/pull} are used for rendezvous-style bulk data transfer over MMIO, and we expand on them in the next section. Lastly, as illustrated in \autoref{figure:user-vm:overview}, the kernel has a dedicated thread that polls for incoming messages from the \systemVm and delivers them to the corresponding user threads.

\subsection{\smallVMM}
\label{subsection:user-vm:vmm}

When designing the \smallVMM, we eschew ABI compatibility with arbitrary OSs and kernels in favor of specializing for the \smallKernel. In the process, we make a series of design choices that depart from traditional VMMs such as Firecracker or CloudHypervisor.

\myparagraph{Shared control-page contract} The \userVm{}'s \smallVMM does not implement standardized virtualization interfaces such as VirtIO~\cite{Virtio:2026} or ACPI~\cite{ACPI:2026} because it only interacts with the \systemVm and is designed for ephemeral execution. Instead, the guest \smallKernel and the \smallVMM use shared memory pages at fixed guest physical addresses~(GPAs) to synchronize control variables: interrupt requests~(IRQs), credit counters, paravirtualized clock, among others.

\myparagraph{PMIO-defined ABI} The \smallVMM relies on PMIO for basic standard I/O from/to the \smallKernel as well as to send commands from the \smallVMM to the \smallKernel, or remote kernel call requests from the \smallKernel to the \smallVMM. This ABI is kept minimal and simple.

\myparagraph{Credit-based flow control} As illustrated in \autoref{figure:user-vm:overview}, the \smallVMM implements a credit-based control flow to minimize the \vm exits issued by the guest when polling for messages. The credit counter is an atomic variable in the guest \smallKernel{}'s control page available to the \smallVMM via shared memory. After incrementing the credit counter, the \smallVMM may also inject an interrupt if the guest is in a halted state.

\myparagraph{Near zero-copy bulk data transfer} To optimize the transfer of large payloads across the \vm boundary, the \smallVMM and the \smallKernel implement a rendezvous-style bulk data transfer protocol over MMIO. As part of the protocol, triggered by the \code{push} or \code{pull} kernel calls, a control message is sent over PMIO with metadata on the GPAs to read/write data from/to with a scatter/gather syntax. The \smallVMM copies the data once from/to the GPAs to/from the transport buffers to the \systemVm, and we expand on ways to eliminate this copy in \autoref{subsection:system-vm:transport}.

\section{\SystemVm}
\label{section:system-vm}

The \systemVm executes inter-kernel calls from \userVms belonging to the same tenant. Next, we present an overview of the \systemVm~(\autoref{subsection:system-vm:overview}), how it isolates requests from different \userVms~(\autoref{subsection:system-vm:isolation}), and its transport layer~(\autoref{subsection:system-vm:transport}).

\subsection{Overview}
\label{subsection:system-vm:overview}

The \systemVm multiplexes inter-kernel calls from different \userVms and translates them to system calls executed inside its \bigKernel. Most importantly, the \systemVm does not contain nor execute any tenant-provided application code, and its image is generic and shared across all tenants. This means that \systemVms, in spite of running a full-weight kernel, can be easily pre-warmed and pooled, thus eliminating the overheads of initializing the different I/O subsystems or device drivers.

\begin{figure}[t]
    \centering
    \includegraphics[width=.9\linewidth]{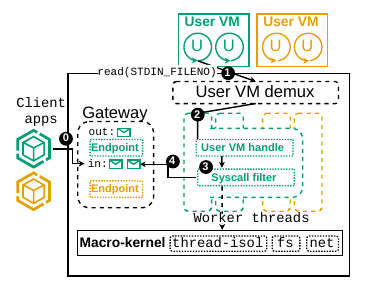}
    \caption{\textbf{\SystemVm architecture} (The \systemVm executes inter-kernel calls from \userVms belonging to the same tenant and interfaces with the different client applications.)}
    \label{figure:system-vm:overview}
\end{figure}

\autoref{figure:system-vm:overview} presents an overview of the \systemVm architecture. The \systemVm maintains a 1:1 mapping between User threads in the \smallKernel~(see \autoref{figure:user-vm:overview}) and worker threads in the \bigKernel: each system call issued by a specific user thread will always be executed by the same worker thread in the \bigKernel. This design simplifies reasoning about kernel call ordering in multi-threaded applications.

\subsection{Isolation}
\label{subsection:system-vm:isolation}

\autoref{figure:system-vm:overview} illustrates the process of executing a \code{read} POSIX call from the perspective of the \systemVm and is complementary to \autoref{figure:user-vm:overview}. Client applications (\eg an interactive user or an event driven workflow) can interact with the \userVm via a gateway TCP port exposed in the \systemVm~(\autoref{figure:system-vm:overview}, \myColourCirc{black}{white}{0}). When the \systemVm receives the read request from the \userVm, it forwards it to the corresponding worker thread based on the \userVm and user thread identifier~\myColourCirc{black}{white}{1}.

Worker threads belonging to the same \userVm can be grouped together for performance isolation using some form of thread isolation (\eg\xspace \code{cgroups-v2} in Linux) and CPU and NUMA affinity. In terms of security isolation, different threads execute in the same virtual address space but each thread executes a very simple logic: it acquires a handle to the \userVm state~\myColourCirc{black}{white}{2}, including the connection to the endpoint, applies the system call filter~\myColourCirc{black}{white}{3} and calls the corresponding \bigKernel system call. This simple datapath in userspace and our use of Rust limit the impact of safety bugs.

After filtering, inter-kernel system calls will be served by the \bigKernel. One notable exception are reads and writes to \code{stdin} and \code{stdout}. We decide to connect the \userVm's standard I/O devices to its gateway endpoint such that client applications can directly interact with the \userVm~\myColourCirc{black}{white}{4}.

\subsection{Transport}
\label{subsection:system-vm:transport}

\sys is a message-oriented \multiKernel. In this regard, \sys takes inspiration from micro-kernel systems~\cite{L3:SOSP:1993,MettEagle:OSDI:2025} and can consequently suffer from similar issues in terms of latencies introduced by the transport layer. For reliable control-plane message delivery, the connections between client applications and \userVms to the \systemVm are established over regular TCP sockets.

The latency of a single TCP message, particularly within the same physical server or in the same server rack, can be well within 40~us~\cite{Demikernel:SOSP:2021} and is unlikely to dominate start up latency or execution time for serverless functions that do little I/O. For large payloads chunked in multiple messages, the bulk data transfer optimization introduced in \autoref{subsection:user-vm:vmm} prevents from sending many small TCP messages, improving throughput.

The same rendezvous-style communication pattern could be extended from the \userVm{}'s \smallVMM to the \bigKernel by setting up a shared memory area between the \bigKernel and the \smallVMM, and exposing it to the former via a mechanism such as \code{ivshmem}~\cite{ivshmem:2026}. Note that the rendezvous communication blocks the sender until the receiver has acknowledged the reception of the message. This introduces safety and starvation issues in micro-kernels where the communication parties are two, potentially distrusting, application processes. \sys sidesteps these issues because in all communications one of the parties is the \userVm{}'s \smallVMM or the \bigKernel, which are part of the system's trusted computing base.

\section{Implementation}
\label{section:implementation}

\sys is implemented from scratch in Rust and made available as open-source at: \url{https://github.com/nanvix/nanvix}.

\myparagraph{\UserVm} The \smallKernel and the \smallVMM are implemented in 17 and 11 kLOC, respectively. The \smallVMM has support for the KVM hypervisor and preliminary support for the Windows Hypervisor Platform~(WHP). \sys can also integrate with existing embedded VMMs as long as they implement the custom PMIO ABI. In fact, \sys can integrate with the Hyperlight VMM~\cite{Hyperlight:2024}, a production-grade alternative to Virtines~\cite{Virtines:EuroSys:2022},  but this integration does not yet have feature-parity with the native \smallVMM.

\myparagraph{\SystemVm} The \systemVm, \ie the components in \autoref{figure:system-vm:overview} besides the \bigKernel, is implemented in 8~kLOC. We use the x86-64 Linux kernel as \bigKernel with a lean configuration profile. We enable KVM and paravirtualized execution, VirtIO, ACPI CPU and memory hotplug, and the core filesystem, namespace, cgroup and memory management features. We disable general-purpose hardware drivers (\eg USB), but consider adding specialized I/O drivers if necessary.
We use Cloud Hypervisor~\cite{CloudHypervisor:2026} as the VMM for the \bigKernel. The \systemVm is always spawned from a snapshot loaded in main memory with on-demand paging. We had to patch CloudHypervisor to load the snapshot lazily, as we found that the upstream version did not have support for this feature crucial to achieve high  density and low start-up times.

\myparagraph{Control plane} We implement a control-plane daemon that manages the life cycle of the different \systemVm and \userVm instances running on the same physical server in 6~kLOC. This component determines the connections between \userVms and \systemVms, as well as the isolation among components from different tenants.

\begin{figure}[t]
    \centering
    \includegraphics[width=.9\linewidth]{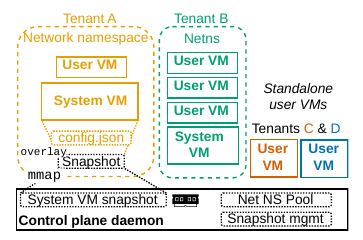}
    \caption{\textbf{\sys deployment} (Groups of \userVms and a \systemVm belonging to the same tenant are deployed inside a network namespace, and \userVms that do not require any I/O can be deployed in standalone mode.)}
    \label{figure:implementation:deployment}
\end{figure}

We illustrate a complete deployment of \sys in \autoref{figure:implementation:deployment}. Groups of \userVms and a \systemVm belonging to the same tenant are deployed inside a network namespace. All \systemVm instances are restored from the same base (memory) snapshot, but we apply an overlay on the snapshot configuration file to introduce per-instance modifications like the console file path or TAP device configuration. For \userVms that do not require a \systemVm, \sys supports deploying them in standalone mode.

\UserVms in \sys are ephemeral and run to completion. Once a \userVm has finished executing the application code it is deleted. \UserVm cold start times are low enough~(see \autoref{table:evaluation:micro:cold-start}) that \sys can afford to not use \code{keep-alive} policies and spawn sandboxes on demand, avoiding the complexities that arise with (safe) sandbox reuse.

\myparagraph{Application compatibility}
\sys can natively run C, C++, and Rust applications after compilation with our toolchain. We ship a GCC-based compilation toolchain with a port of \code{newlib} and a POSIX shim layer. For Rust applications, we add a custom \sys target. \sys' also has ports of Wasmtime, CPython and Deno runtimes to execute WASM bytecode, Python, or JS programs, respectively.

\myparagraph{Future work}
\sys is an early-stage prototype and, as such, has limitations that we plan on addressing in the future. The current \userVm prototype only implements the \code{i686} 32-bit architecture and only supports single-core execution\footnote{Work on \code{x86\_64} 64-bit has already started.}. The shared memory, zero-copy, datapath between the \userVm and the \systemVm is also not yet merged. In terms of application compatibility, currently Rust applications targeting \sys{} have increasing support for \code{std}. There is also on-going effort to support the NodeJS runtime. None of the above are inherent limitations of \sys' \multiKernel design, and are a matter of engineering effort.

Acknowledging the above limitations, \sys{} codebase is actively maintained and is undergoing formal verification with Verus~\cite{Verus:SOSP:2024}. The system is developed and tested in \textit{company X} and will soon be deployed in production.

\section{Evaluation}
\label{section:evaluation}

In the evaluation we aim to answer the following questions:
\begin{enumerate}[label=\roman*]
    \item What is the performance of \sys and how does it compare to state-of-the-art serverless systems?~(\autoref{subsection:evaluation:micro})
    \item What are the advantages of deploying \sys in a serverless fleet when replaying a production trace?~(\autoref{subsection:evaluation:end-to-end})
\end{enumerate}

\subsection{Experimental set-up}
\label{subsection:evaluation:set-up}

\myparagraph{Baselines} We compare \sys (\sysShort) against the open-source solutions used by the leading serverless providers: Firecracker \code{v1.12.0} (FC), CloudHypervisor \code{v45.0} (CLH), and gVisor \code{v20250505} (gV). We also compare against a production unikernel, Unikraft~\cite{Unikraft:EuroSys:2021} \code{v0.12.3} (UK), as it is increasingly being adopted in the context of serverless~\cite{SEUSS:EuroSys:2020,SURE:SoCC:2024}, and against Linux processes (Proc) and Hyperlight (HL)~\cite{Hyperlight:2024} as lower-bounds on the overhead to instantiate any application and any \vm, respectively.

\myparagraph{Hardware} We run our experiments on a bare-metal Intel Xeon server with two sockets and ten cores per socket. We disable hyperthreading and frequency scaling. Our server has 32 GiB of memory, two NUMA regions, and runs Ubuntu 24.04 with the \code{6.8.0-90-generic} kernel.

\subsection{Microbenchmarks}
\label{subsection:evaluation:micro}

To understand the benefits and overheads of \sys{}' \multiKernel \os architecture, we deployed a HTTP echo server written in Rust that sends and receives a 32~B payload, and measure different key performance metrics.

\myparagraph{Cold start latency} We first measure cold start latency, \ie the time elapsed between sending the HTTP echo request and receiving the response when there are no previously-running instances. The cold start latency includes the time to provision the execution environment, its network devices (excluding the network namespace), and the application itself. We include two additional baselines that benefit from having executed the function once: Firecracker (FC-S) and CloudHypervisor (CLH-S) starting the function from a generic \os snapshot. For \sys, the cold start latency measures the time to spawn a \systemVm and a \userVm, whereas \sys{}-uV measures the time to spawn only a \userVm. We report the p50 and p99 cold start latencies over 1000 iterations.

\begin{table}[t]
    \centering
    \small
    \begin{tabular}{p{2.6cm} p{1.3cm} p{1.3cm} p{1.6cm}}
    \toprule
        \multicolumn{1}{c}{\textbf{Baseline}} & \multicolumn{1}{c}{\textbf{p50 (ms)}} & \multicolumn{1}{c}{\textbf{p99 (ms)}} & \multicolumn{1}{c}{\textbf{Slowdown}} \\
    \midrule
        \multicolumn{4}{c}{\textit{First time}} \\
    \midrule
        Firecracker & 1055.05 & 1097.50 & 37.76x \\
        Cloud Hypervisor & 1047.55 & 1069.77 & 37.49x \\
        Unikraft & 321.28 & 347.79 & 11.50x \\
        gVisor & 161.10 & 1185.52 & 5.77x \\
        \textbf{\sys{}} & \textbf{27.94} & \textbf{41.29} & \textbf{1.00x} \\
        Hyperlight & 7.68 & 22.41 & 0.27x \\
        Process & 3.32 & 18.82 & 0.12x \\
    \midrule
        \multicolumn{4}{c}{\textit{Known function}} \\
    \midrule
        CloudHypervisor-S & 41.44 & 118.23 & 5.71x \\
        Firecracker-S & 69.45 & 83.77 & 9.56x \\
        \textbf{\sys{}-uV} & \textbf{7.26} & \textbf{11.37} & \textbf{1.00x} \\
    \bottomrule
    \end{tabular}
    \caption{\textbf{Cold start latency} (The \textit{First time} section reports slowdown relative to \sys{}, spawning both \userVm and \systemVm, while the \textit{Known function} section reports slowdown relative to \sys{} spawning only a \userVm.)}
    \label{table:evaluation:micro:cold-start}
\end{table}

\autoref{table:evaluation:micro:cold-start} summarizes our results. \sys has two orders of magnitude lower cold start latencies than systems that boot Linux from scratch (\ie Firecracker and CloudHypervisor), and 6-11$\times$ lower cold start latencies compared to specialized solutions such as gVisor and Unikraft. In both cases, a noticeable part of the difference comes from setting up the network, including TAP devices. For Unikraft, we use upstream QEMU \code{v8.2.2} as the VMM, whereas the original paper used a highly modified one~\cite{Unikraft:EuroSys:2021} and only reported boot time, not time to serve the first HTTP request.

Compared to snapshot-based baselines (\ie Firecracker-S and CloudHypervisor-S), \sys{} shows similar cold start latency. This is because, under the hood, \sys{} spawns the \systemVm from a snapshot using CloudHypervisor, so the latter's cold start latency dominates end-to-end latency and determines its p99. The reason why \sys{} is 10~ms faster than CloudHypervisor-S is that the \systemVm uses an in-memory filesystem whereas CLH-S uses a block device. The reason why CloudHypervisor-based systems are marginally faster than Firecracker-S is that the latter must create the network TAP device in a separate (\code{sudo}) bash script, and just the call to \code{sudo} adds upwards of 10~ms.

Most importantly, if we exclude the time to create the \systemVm because \eg there is another in-flight function from the same tenant, \sys{}-uV has 6-10$\times$ lower cold start latencies than snapshot-based systems, and is comparable to the reference baselines. This makes it possible to spawn a new \userVm for each new request.

\myparagraph{\UserVm boot time breakdown} \autoref{table:evaluation:micro:cold-start} presents the fine-grain cold start latencies in \sys. When the \systemVm is already running, a \userVm starts in 7~ms at p50, and 11~ms at p99. This time includes sending an HTTP request to the control-plane, spawning a new \userVm, opening a port in the \systemVm{}'s gateway, reading the payload from the user, sending it all the way to the user thread in the \userVm, and back.

\begin{table}[t]
    \centering
    \begin{tabular}{lrrr}
        \toprule
            \textbf{Phase} & \textbf{p50 (us)} & \textbf{p95 (us)} & \textbf{p99 (us)} \\
        \midrule
            \code{channel\_setup}   & 12   & 18   & 61   \\
            \code{partition\_create}& 470  & 613  & 1131 \\
            \code{vmem\_create}     & 49   & 77   & 92   \\
            \code{vcpu\_create}     & 158  & 251  & 275  \\
            \code{kernel\_load}     & 1636 & 3155 & 3339 \\
            \code{initrd\_load}     & 60   & 112  & 131  \\
            \code{vcpu\_reset}      & 19   & 36   & 39   \\
            \code{thread\_spawn}    & 38   & 61   & 66   \\
            \code{guest\_exec}      & 952  & 1415 & 1751 \\
            \code{exit\_handling}   & 47   & 64   & 80   \\
        \midrule
            \textbf{Total}            & 3553 & 5069 & 5460 \\
        \bottomrule
    \end{tabular}

    \caption{\textbf{\UserVm boot time breakdown} (We measure the latency of each phase involved in booting a \userVm and executing the HTTP echo application in standalone mode, \ie without a \systemVm and interacting with the \userVm via standard I/O.)}
    \label{table:evaluation:micro:boot-time-breakdown}
\end{table}

To isolate the time to create a \userVm, \autoref{table:evaluation:micro:boot-time-breakdown} presents a breakdown of the time spent creating (and executing) the \userVm for the same application in standalone mode, where we communicate with the \userVm via standard I/O (see \autoref{figure:implementation:deployment}). Removing the control plane and \systemVm interactions brings cold start latencies to 3.5~ms at p50 and 5.5~ms at p99. This latency is dominated by the time to load the 1.3~MiB kernel image into the guest's address space. The application binary itself, presented as an \code{initrd}, only occupies 45~KiB and loads in negligible time. The second contributor to latency is the guest execution time, which includes all the time the vCPU is executing in the guest. These latencies are comparable to other embedded VMMs~\cite{Hyperlight:2024,Virtines:EuroSys:2022}.

\myparagraph{Memory footprint} To understand the relationship between cold start latency and memory consumption, we next deploy instances in a closed-loop until we consume 1~GiB of system memory. We present the cold start latency for each instance and their memory contribution, calculated by subtracting the before and after readings of \code{MemAvailable} in \code{/proc/meminfo}, in a scatter plot. We present the p50 from 10 iterations across both and the minimum number of instances deployed in 1~GiB across all iterations. For \sys, we consider two ends of the spectrum in terms of deployment: only one \systemVm (\sysShort) as a best case scenario, or one \systemVm per \userVm (\sysShort-1to1) as a worst case scenario.

\begin{figure*}[t]
    \centering
    \begin{subfigure}{.48\linewidth}
        \centering
        \includegraphics[width=\linewidth]{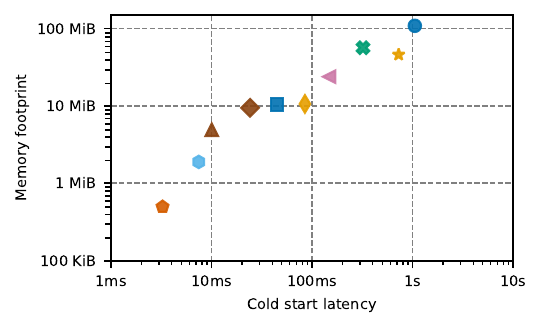}
        \caption{Cold start against memory footprint}
        \label{figure:evaluation:micro:density:mem-footprint}
    \end{subfigure}
    ~
    \begin{subfigure}{.48\linewidth}
        \centering
        \includegraphics[width=\linewidth]{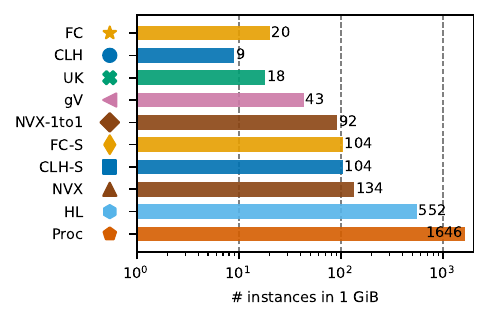}
        \caption{Instance count}
        \label{figure:evaluation:micro:density:count}
    \end{subfigure}

    \caption{\textbf{Deployment density} (We spawn sandboxes in a closed-loop until we consume 1 GiB of system memory, and measure each sandbox's contribution by reading \texttt{MemAvailable} from \texttt{/proc/meminfo}. Each dot represents the median cold start latency and memory footprint across runs. The sandbox count is the minimum number of sandboxes, across all 10 runs, we can fit in 1 GiB.)}
    \label{figure:evaluation:micro:density}
\end{figure*}

\autoref{figure:evaluation:micro:density:mem-footprint} summarizes our results. The cold start latencies, on the X axis, are comparable to those reported in \autoref{table:evaluation:micro:cold-start}. In terms of the memory footprint, the same application executing in a \userVm has an order of magnitude lower memory footprint than executing in a (micro) Linux \vm, and a 30-50\% lower memory footprint than the same application being restored from a snapshot with on-demand paging. This is due to \sys native VMM being much smaller and simpler than CloudHypervisor or Firecracker, and because \sys does not virtualize devices via VirtIO. \sysShort-1to1 is on par with CLH-S because the \systemVm instance dominates the memory footprint.

These reductions in memory footprint are reflected in higher instance counts per 1~GiB of memory, which we present in \autoref{figure:evaluation:micro:density:count}. The ratios are not exactly the same because the instance counts are influenced by peak memory consumption of outliers, whereas the scatter plot shows medians. Compared to the reference baselines, isolating sandboxes in minimal \vms (\ie HL) reduces deployment count by 3.5$\times$ compared to Proc, and adding a minimal \os to the \vm (\ie \sysShort) reduces it by a similar factor.

Most importantly, \sys in a worst-case adversarial scenario where each \userVm instance is connected to a different \systemVm (\ie \sysShort-1to1) is on par with snapshot-based baselines. This situation corresponds to a deployment where each tenant only runs a single application once, and is extremely rare~(see \autoref{figure:motivation:amortization}). On the other hand, this experiment is a best-case scenario for snapshot based baselines because we run the same application for all requests, and thus all requests can reuse the exact same snapshot.

\begin{figure*}[t]
    \centering
    \begin{minipage}[t]{0.48\textwidth}
        \vspace{0pt}
\definecolor{latFirecracker}{RGB}{230,159,0}
\definecolor{latCloudHypervisor}{RGB}{0,114,178}
\definecolor{latUnikraft}{RGB}{0,158,115}
\definecolor{latgVisor}{RGB}{204,121,167}
\definecolor{latNanvix}{RGB}{139,69,19}
\definecolor{latHyperlight}{RGB}{86,180,233}
\definecolor{latProcess}{RGB}{213,94,0}
    \centering
    \small
    \captionsetup{type=table}
    \begin{tabular}{p{2.9cm} c c c}
    \toprule
        \multicolumn{1}{c}{\textbf{Baseline}} & \textbf{p50 (us)} & \textbf{p99 (us)} & \textbf{Slowdown} \\
    \midrule
        \tikz[baseline=-0.6ex, x=1ex, y=1ex]{\filldraw[draw=latFirecracker, fill=latFirecracker] (0,1.0ex) -- (0.24ex,0.31ex) -- (0.95ex,0.31ex) -- (0.38ex,-0.12ex) -- (0.59ex,-0.81ex) -- (0,-0.38ex) -- (-0.59ex,-0.81ex) -- (-0.38ex,-0.12ex) -- (-0.95ex,0.31ex) -- (-0.24ex,0.31ex) -- cycle;}\hspace{0.45em}Firecracker & 240 & 345 & 0.74x \\
        \tikz[baseline=-0.6ex, x=1ex, y=1ex]{\filldraw[draw=latCloudHypervisor, fill=latCloudHypervisor] (0,0) circle (0.75ex);}\hspace{0.45em}Cloud Hypervisor & 195 & 262 & 0.60x \\
        \tikz[baseline=-0.6ex, x=1ex, y=1ex]{\draw[draw=latUnikraft, line width=0.45pt] (-0.8ex,-0.8ex) -- (0.8ex,0.8ex) (-0.8ex,0.8ex) -- (0.8ex,-0.8ex);}\hspace{0.45em}Unikraft & 214 & 271 & 0.66x \\
        \tikz[baseline=-0.6ex, x=1ex, y=1ex]{\filldraw[draw=latgVisor, fill=latgVisor] (-0.95ex,0) -- (0.7ex,0.9ex) -- (0.7ex,-0.9ex) -- cycle;}\hspace{0.45em}gVisor & 387 & 495 & 1.19x \\
        \textbf{\tikz[baseline=-0.6ex, x=1ex, y=1ex]{\filldraw[draw=latNanvix, fill=latNanvix] (0,0.9ex) -- (-0.9ex,-0.7ex) -- (0.9ex,-0.7ex) -- cycle;}\hspace{0.45em}\sys{}} & \textbf{325} & \textbf{422} & \textbf{1.00x} \\
        \tikz[baseline=-0.6ex, x=1ex, y=1ex]{\filldraw[draw=latHyperlight, fill=latHyperlight] (-0.9ex,0) -- (-0.45ex,0.8ex) -- (0.45ex,0.8ex) -- (0.9ex,0) -- (0.45ex,-0.8ex) -- (-0.45ex,-0.8ex) -- cycle;}\hspace{0.45em}Hyperlight & 207 & 274 & 0.64x \\
        \tikz[baseline=-0.6ex, x=1ex, y=1ex]{\filldraw[draw=latProcess, fill=latProcess] (0,1.0ex) -- (-0.95ex,0.3ex) -- (-0.6ex,-0.9ex) -- (0.6ex,-0.9ex) -- (0.95ex,0.3ex) -- cycle;}\hspace{0.45em}Process & 121 & 174 & 0.37x \\
    \bottomrule
    \end{tabular}
    \caption{\textbf{HTTP round-trip latency} (Slowdown is reported relative to \sys{}.)}
    \label{table:evaluation:micro:round-trip}

    \end{minipage}
    \hfill
    \begin{minipage}[t]{0.48\textwidth}
        \vspace{-11pt}
        \centering
        \includegraphics[width=.9\linewidth]{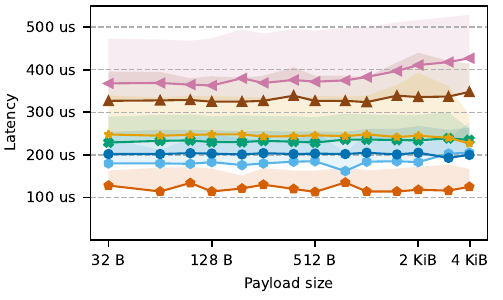}
        \vspace{-11pt}
        \caption{\textbf{Throughput characteristic} (We present the p50, marker, and p99, faded. See \autoref{table:evaluation:micro:round-trip} for the legend.)}
        \label{figure:evaluation:micro:throughput}
    \end{minipage}
\end{figure*}

\myparagraph{Warm start latency} We have shown that \sys' \multiKernel design reduces cold start latency and improves memory density, but it also introduces an additional hop in the data path (see \autoref{figure:system-vm:overview}). \autoref{table:evaluation:micro:round-trip} presents the round-trip latency of sending an HTTP echo request with a 32 B payload, and we report p50 and p99 across a million iterations and the slowdown compared to \sys{}.

The results in \autoref{table:evaluation:micro:round-trip} show that \sys introduces, at worst, a 50\% overhead on round-trip latency compared to the other virtualized baselines. Our profiling indicates that this overhead is not due to the additional hop, but rather to the additional signaling in the \userVm. This could be addressed by exposing a shared memory area from the \systemVm to the \userVm, and is something we plan on exploring as future work.

In \autoref{figure:evaluation:micro:throughput} we repeat the same experiment increasing the size of the payload. Thanks to the bulk data transfer optimization described in \autoref{subsection:user-vm:vmm}, increasing the size of the payload does not affect request latency in \sys{} even though the payload size exceeds the PMIO message size. However, this is not the case when the payload size exceeds a guest page size of 4 KiB. The current bulk data transfer protocol implementation only supports transferring data within a page, but we are working on an improved version that uses scatter/gather syntax across guest pages.

\subsection{End-to-end macrobenchmark}
\label{subsection:evaluation:end-to-end}

\begin{figure*}[t]
    \centering
    \begin{minipage}[t]{0.48\textwidth}
        \vspace{0pt}
        \centering
        \includegraphics[width=\linewidth]{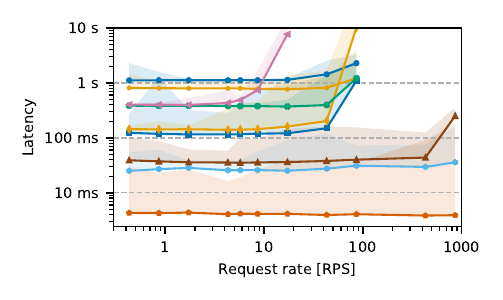}
        \vspace{-28pt}
        \caption{\textbf{Trace replay} (We replay the Huawei trace and measure the per-request p50, marker, and p99, shade, latency. See the table at the right-hand side for the legend.)}
        \label{figure:evaluation:macro:trace}
    \end{minipage}
    \hfill
    \begin{minipage}[t]{0.48\textwidth}
        \vspace{0pt}
\definecolor{trFirecracker}{RGB}{230,159,0}
\definecolor{trCloudHypervisor}{RGB}{0,114,178}
\definecolor{trUnikraft}{RGB}{0,158,115}
\definecolor{trgVisor}{RGB}{204,121,167}
\definecolor{trFirecrackerSnapshot}{RGB}{230,159,0}
\definecolor{trCloudHypervisorSnapshot}{RGB}{0,114,178}
\definecolor{trHyperlight}{RGB}{86,180,233}
\definecolor{trProcess}{RGB}{213,94,0}
\definecolor{trNanvix}{RGB}{139,69,19}
    \centering
    \small
    \setlength{\tabcolsep}{4pt}
    \captionsetup{type=table}
    \begin{tabular}{lrrrr}
    \toprule
        \textbf{Baseline} & \textbf{Peak RPS} & \textbf{Mem (MiB)} & \multicolumn{2}{l}{\textbf{\# Servers}} \\
\midrule \hline
        \tikz[baseline=-0.6ex, x=1ex, y=1ex]{\filldraw[draw=trFirecracker, fill=trFirecracker] (0,1.0ex) -- (0.24ex,0.31ex) -- (0.95ex,0.31ex) -- (0.38ex,-0.12ex) -- (0.59ex,-0.81ex) -- (0,-0.38ex) -- (-0.59ex,-0.81ex) -- (-0.38ex,-0.12ex) -- (-0.95ex,0.31ex) -- (-0.24ex,0.31ex) -- cycle;}\hspace{0.45em}Firecracker & 43.1 & 2211 & 100 & 20x \\
        \tikz[baseline=-0.6ex, x=1ex, y=1ex]{\filldraw[draw=trCloudHypervisor, fill=trCloudHypervisor] (0,0) circle (0.75ex);}\hspace{0.45em}Cloud Hypervisor & 43.1 & 5479 & 100 & 20x \\
        \tikz[baseline=-0.6ex, x=1ex, y=1ex]{\draw[draw=trUnikraft, line width=0.45pt] (-0.8ex,-0.8ex) -- (0.8ex,0.8ex) (-0.8ex,0.8ex) -- (0.8ex,-0.8ex);}\hspace{0.45em}Unikraft & 43.1 & 1617 & 100 & 20x \\
        \tikz[baseline=-0.6ex, x=1ex, y=1ex]{\filldraw[draw=trgVisor, fill=trgVisor] (-0.95ex,0) -- (0.7ex,0.9ex) -- (0.7ex,-0.9ex) -- cycle;}\hspace{0.45em}gVisor & 8.62 & 151 & 500 & 100x \\
        \tikz[baseline=-0.6ex, x=1ex, y=1ex]{\filldraw[draw=trFirecrackerSnapshot, fill=trFirecrackerSnapshot] (0,1.0ex) -- (-0.9ex,0) -- (0,-1.0ex) -- (0.9ex,0) -- cycle;}\hspace{0.45em}Firecracker-S & 43.1 & 382 & 100 & 20x \\
        \tikz[baseline=-0.6ex, x=1ex, y=1ex]{\filldraw[draw=trCloudHypervisorSnapshot, fill=trCloudHypervisorSnapshot] (-0.75ex,-0.75ex) rectangle (0.75ex,0.75ex);}\hspace{0.45em}CloudHypervisor-S & 43.1 & 125 & 100 & 20x \\
        \tikz[baseline=-0.6ex, x=1ex, y=1ex]{\filldraw[draw=trHyperlight, fill=trHyperlight] (-0.9ex,0) -- (-0.45ex,0.8ex) -- (0.45ex,0.8ex) -- (0.9ex,0) -- (0.45ex,-0.8ex) -- (-0.45ex,-0.8ex) -- cycle;}\hspace{0.45em}Hyperlight & 862 & 230 & 5 & 1x \\
        \tikz[baseline=-0.6ex, x=1ex, y=1ex]{\filldraw[draw=trProcess, fill=trProcess] (0,1.0ex) -- (-0.95ex,0.3ex) -- (-0.6ex,-0.9ex) -- (0.6ex,-0.9ex) -- (0.95ex,0.3ex) -- cycle;}\hspace{0.45em}Process & 4310 & 553 & 1 & 0.20x \\
    \midrule
        \textbf{\tikz[baseline=-0.6ex, x=1ex, y=1ex]{\filldraw[draw=trNanvix, fill=trNanvix] (0,0.9ex) -- (-0.9ex,-0.7ex) -- (0.9ex,-0.7ex) -- cycle;}\hspace{0.45em}\sys{}} & \textbf{862} & \textbf{1087} & \textbf{5} & \textbf{1x} \\
    \bottomrule
    \end{tabular}
    \caption{\textbf{Trace replay summary} (Peak RPS is the maximum sustainable request rate, and peak memory is taken from the corresponding run. The number of servers is projected based on the single-node peak RPS and the trace size.)}
    \label{table:evaluation:trace-capacity}

    \end{minipage}
    \vspace{-5pt}
\end{figure*}

In this section we show how \sys{}' improvements in cold start latency and memory density influence fleet-wide utilization metrics when replaying a production serverless trace. We replay the first minute of the first region of the Huawei trace from 2025~\cite{ServerlessColdStartsAndWhereToFindThem:EuroSys:2025} because it includes the unique tenant identifier for each function invocation. We replay a downsampled version of the trace, and keep reducing the downsampling ratio, \ie increasing requests-per-second~(RPS), until we either exhaust memory capacity or measure a spike in per-request latency. Based on the maximum sustainable throughput on a single node we derive how many servers we would need to run this trace assuming perfect load balancing.

We drive the trace execution from an open-loop client with a bounded number of in-flight requests, but we size this upper-bound such that the client is never saturated. For \sys{}, the control plane handles the invocation requests and spawns \systemVm and \userVm instances accordingly. For the other baselines, we implement a multi-threaded HTTP server in Rust with a worker pool that we are also careful not to saturate. Each request executes the same HTTP echo application and each request experiences a cold start. This is a slightly pathological case, \ie one where the serverless platform is configured with a \code{keep-alive} of 0, but it helps us test the limits of each solution.

\autoref{figure:evaluation:macro:trace} presents the p50 and p99 of the per-request latency as we decrease the downsampling factor of the trace replay, thus increasing the overall RPS. The trace has sudden bursts of requests, very common in serverless traces~\cite{ServerlessInTheWild:ATC:2020}, which introduce higher per-request latency. Still, \sys{} has consistently one to two orders of magnitude lower per-request latency, and achieves an order-of-magnitude improvement in maximum sustainable RPS.

\autoref{table:evaluation:trace-capacity} summarizes the peak RPS achieved by each system, as well as the peak memory consumed on the corresponding replay. Based on the memory consumption we can assert that none of the systems exhaust the system memory, so in this case the determining factor for peak RPS is contention on host resources. Given that \sys{} has a simple per-request virtual environment and multiplexes all I/O requests from the same tenant, it can achieve much higher peak RPS. Note that network namespace creation and allocation can very quickly dominate request latency, much earlier than the current system's elbow points~\cite{RunD-V:TOCS:2025,AFaaS:OSDI:2025}. To prevent namespace allocation from influencing the results, both \sys{}' control plane and the other baseline's server allocate namespaces from a pre-initialized pool.

Most importantly, based on the peak single-node RPS, a serverless provider using \sys{} would need 20-100$\times$ less physical servers to serve the same workload. This is a substantial improvement in achieved density and justifies \sys{}' design. This projection of total number of servers assumes perfect load-balancing and overlooks the additional challenges from routing functions to a server with a \systemVm from the same tenant. We also keep \systemVms alive after execution, but do not pre-warm them in a pool (which could be achievable in practice). Given the substantial improvements achieved by \sys{} we do not believe these simplifications affect our claims.


\section{Related Work}
\label{section:related-work}

\myparagraph{Execution environments for serverless applications}
Serverless providers navigate a trade-off between security, performance, density, and compatibility when choosing an execution environment for applications. Systems like Faasm~\cite{Faasm:ATC:2020} or CloudFlare Workers~\cite{CloudFlareWorkers:2026} achieve low cold starts and high density sacrificing security and compatibility by executing applications in WebAssembly sandboxes. Container-based solutions offer compatibility and performance similar to solutions based on system call filtering like gVisor~\cite{gVisor:HotCloud:2019} or Junction~\cite{Junction:NSDI:2024}. In both cases, applications still share a common \os kernel, posing a security threat.

The gold standard for security in the cloud is hypervisor isolation as provided by \vms. \vms, however, are slow to start up and have a high memory footprint. Even micro-\vms like Firecracker~\cite{Firecracker:NSDI:2020} or Cloud Hypervisor~\cite{CloudHypervisor:2026} introduce substantial overheads in terms of cold start latency and deployment density. Systems like Virtines~\cite{Virtines:EuroSys:2022} and Hyperlight~\cite{Hyperlight:2024} showed that these overheads are not due to hypervisor isolation, but rather expensive initialization of \os subsystems.

Specialized \os solutions, like unikernels~\cite{Unikraft:EuroSys:2021,SURE:SoCC:2024,SEUSS:EuroSys:2020}, try and minimize the guest \os by linking it with the application code in a single binary, but still introduce latency and memory overheads which are an open problem addressed by related work~\cite{MemoryMatters:SoCC:2025}. In addition, most unikernels have compatibility issues due to their single-address space design, that prevents them from supporting applications that use the \code{fork} system call~\cite{uFork:SOSP:2025}. If a bug or vulnerability is found in the unikernel's system libraries, \emph{all} affected applications must be recompiled, complicating the management and deployment of such systems at scale.

\myparagraph{Intra-tenant resource sharing}
Ultimately, execution environments adopting \vms and full-weight OSs for compatibility will compromise either cold start latency and/or deployment density. As described in \autoref{section:motivation}, these limitations can be addressed by sharing components across applications belonging to the same tenant. Some systems like RunD-V~\cite{RunD-V:TOCS:2025} share the \vm and \os explicitly by spawning applications inside containers in a per-tenant \vm, whereas others like Ant FaaS~\cite{AFaaS:OSDI:2025} or AWS Lambda Snap Start~\cite{AwsSnapStart:2026} exploit \vm snapshots to: (i) reduce cold start latency and per-instance memory footprint~\cite{REAP:ASPLOS:2021}, and (ii) reduce memory duplication across instances with generic \os images. As argued in \autoref{section:motivation} all these systems introduce hard-to-quantify trade-offs in resource contention, system complexity, and disk pressure.

\myparagraph{\os disaggregation}
\sys{} follows a different approach in that it disaggregates the components in the guest \os in two virtualized environments. \sys{} \multiKernel design is inspired by other distributed OSs like Barrelfish~\cite{Barrelfish:SOSP:2009} or Helios~\cite{Helios:SOSP:2009}, and disaggregated OSs like LegOS~\cite{LegOS:OSDI:2018} and FractOS~\cite{FractOS:EuroSys:2022}. The \userVm runs a \smallKernel with a message oriented architecture inspired from that of L3~\cite{L3:SOSP:1993} and its multiple successors~\cite{seL4:SOSP:2009}. Dandelion~\cite{Dandelion:SOSP:2025} argues for an explicit I/O and compute separation which enables the system to achieve elasticity at the cost of application compatibility.

\myparagraph{Alternative isolation mechanisms}
Some systems explore different forms of hardware-based isolation and sharing. ORC~\cite{ORC:OSDI:2023} uses hardware capabilities to share memory at object granularity whereas CubicleOS~\cite{CubicleOS:ASPLOS:2021} and ERIM~\cite{ERIM:Security:2019} use compartments and memory protection keys. These systems depart from the traditional \vm abstraction and generic hardware features, and are orthogonal to \sys.


\section{Discussion}
\label{sections:discussion}

We believe \sys{} disaggregated \multiKernel design offers a deployment and configuration flexibility that can be leveraged in many different use-cases where efficient ephemeral execution matters.

\myparagraph{Single-tenant deployments}
We envision deployment scenarios where it is not necessary to execute the \systemVm inside a virtualized environment, and it instead can execute in the host as a \emph{system process}. In this case the system process would be similar to an emulated device.

\myparagraph{Ephemeral execution of untrusted code}
\UserVms, particularly in standalone mode, introduce minimal cold start latencies with strong isolation (see \autoref{table:evaluation:micro:boot-time-breakdown}). \UserVms could therefore be used to sandbox execution of LLM generated code that only needs access to standard I/O.

\myparagraph{Disaggregated deployment}
At the moment, we only consider \sys{} deployments where the \systemVm and \userVm are co-located on the same physical server. However, both \vms already communicate via TCP, so they could also be deployed in separate servers for improved load balancing. Such a set-up would introduce additional latency to inter-kernel I/O requests, so it should be weighed carefully.

\myparagraph{Hardware offload}
The current \sys{} prototype uses in-\bigKernel options for the different I/O stacks in the \systemVm. However, with the increased adoption of hardware offload~\cite{LineFS:SOSP:2021}, kernel bypass~\cite{Demikernel:SOSP:2021,Junction:NSDI:2024}, and storage disaggregation~\cite{AzureRDMA:NSDI:2023}, we envision a \sys{} deployment where different I/O stacks are offloaded to different hardware devices. Whether \userVm requests are still routed through a centralized component or not is an open question that we plan on exploring as future work.

\myparagraph{Lessons learned}
When we started working on \sys{} we assumed deployment density was determined by per-application memory footprint. However, after trace analysis and experimentation we realized that cold start latencies also played an important role and, most importantly, resource contention on the host \os resources.

Choosing to provide application compatibility at the POSIX layer introduces initial porting efforts~\cite{Loupe:ASPLOS:2024} but unlocks new opportunities for optimization. In comparison to other systems that also offer POSIX compatibility like unikernels~\cite{Unikraft:EuroSys:2021} or other libOSs~\cite{Mirage:ASPLOS:2013} that are constrained by their execution in a single-address space, which limits the type of applications they can support, we do not envision any such limitations for \sys{}.


\section{Conclusions}
\label{section:conclusions}

\sys{} revisits the long-standing \vm abstraction and shows that a principled split between ephemeral execution state and persistent system state enables a fundamentally more efficient serverless substrate. By introducing \userVms{} with a minimal, purpose-built \smallKernel and a shared \systemVm{} that encapsulates heavyweight OS functionality, \sys{} avoids the resource stranding, resizing overheads, and snapshot complexity that limit existing container- and \vm-based approaches. This separation allows providers to safely amortize \os and runtime costs across same-tenant applications while preserving strong isolation boundaries and maintaining compatibility with unmodified applications.

Our evaluation demonstrates that this split design translates into tangible gains: \userVms start in a few milliseconds, \systemVms can be pooled and reused across tenants, and the combined deployment path remains well below the latency of hot-plugging or snapshot-heavy baselines. More importantly, \sys{} improves deployment density by making intra-tenant sharing explicit and efficient, enabling significantly more sandboxes per GiB and reducing the number of servers required to serve production-scale traces.

{\footnotesize\bibliographystyle{plain}
    \bibliography{references/articles,references/websites}}

\end{document}